\PassOptionsToPackage{dvipsnames,table}{xcolor}
\documentclass[sigconf]{acmart}
\usepackage{multirow}
\usepackage{multicol}

\usepackage[colorinlistoftodos]{todonotes}

% ...
\AtBeginDocument{%
  \providecommand\BibTeX{{%
    \normalfont B\kern-0.5em{\scshape i\kern-0.25em b}\kern-0.8em\TeX}}}

\setcopyright{acmcopyright}
\copyrightyear{2021} 
\acmYear{2021} 
\setcopyright{acmcopyright}\acmConference[CHI '21]{CHI Conference on Human Factors in Computing Systems}{May 8--13, 2021}{Yokohama, Japan}
\acmBooktitle{CHI Conference on Human Factors in Computing Systems (CHI '21), May 8--13, 2021, Yokohama, Japan}
\acmPrice{15.00}
\acmDOI{10.1145/3411764.3445523}
\acmISBN{978-1-4503-8096-6/21/05}

\begin{document}

\title[Personalised Recommendations in Mental Health Apps]{Personalised Recommendations in Mental Health Apps: The Impact of Autonomy and Data Sharing}

\author{Svenja Pieritz}
\authornote{Both authors contributed equally to this work}
\affiliation{%
\institution{Telefonica Alpha, Spain}
}
\email{svenja.pieritz@gmail.com}

\author{Mohammed Khwaja}
\authornotemark[1]
\authornote{The author is also a part of Koa Health, Spain}
\affiliation{%
\institution{Imperial College London, UK}
}
\email{mohammed.khwaja16@imperial.ac.uk}

\author{A. Aldo Faisal}
\affiliation{%
\institution{Imperial College London, UK}
}
\email{a.faisal@imperial.ac.uk}

\author{Aleksandar Matic}
\affiliation{%
\institution{Koa Health, Spain}
}
\email{aleksandar.matic@koahealth.com}

\renewcommand{\shortauthors}{Pieritz and Khwaja et al}

\begin{abstract}
The recent growth of digital interventions for mental well-being prompts a call-to-arms to explore the delivery of personalised recommendations from a user's perspective. In a randomised placebo study with a two-way factorial design, we analysed the difference between an autonomous user experience as opposed to personalised guidance, with respect to both users’ preference and their actual usage of a mental well-being app. Furthermore, we explored users’ preference in sharing their data for receiving personalised recommendations, by juxtaposing questionnaires and mobile sensor data. Interestingly, self-reported results indicate the preference for personalised guidance, whereas behavioural data suggests that a blend of autonomous choice and recommended activities results in higher engagement. Additionally, although users reported a strong preference of filling out questionnaires instead of sharing their mobile data, the data source did not have any impact on the actual app use. We discuss the implications of our findings and provide takeaways for designers of mental well-being applications.

\end{abstract}

\ccsdesc[500]{Human-centered computing~Human computer interaction (HCI)}
\ccsdesc[300]{Human-centered computing~Empirical studies in ubiquitous and mobile computing}

\keywords{User Perception; Personalisation; Recommender Systems; Personality Traits}

\maketitle

\section{Introduction}

Digital mental well-being interventions present the promise to mitigate the global shortage of mental healthcare professionals in a cost-effective and scalable manner~\cite{tal2017digital}. Their emergence has been accelerated by the experiences of the COVID-19 pandemic~\cite{world2020coronavirus} and its impact on mental health~\cite{pfefferbaum2020mental}. The growth of the digital mental health space has been paralleled by the rapid increase in research and development of new interventions and content. Benefits of rich content are indisputable, yet a vast amount of choices can also misfire—which is known as a paradox of choice~\cite{schwartz2004paradox}. For this reason, we are witnessing the advent of recommender systems also in digital mental health platforms~\cite{khwaja2019aligning}. Although personalised recommendations represent an important aid with respect to the choice overload and moreover in improving the intervention effectiveness, delivering those recommendations entails two main challenges. Firstly, how to balance autonomy and personalised guidance has become an important topic in the design of personalised technologies. Secondly, data sharing concerns are undetachable from the automatic personalisation models. Both challenges have a very specific relevance when it comes to digital health applications~\cite{burr2020ethics}. 

While users' autonomy is one of the common principles in designing digital experiences~\cite{peters2018designing}, patients in traditional doctor-patient settings typically expect (and often prefer) to “be told what to do” rather than to “do what they want”. This raises an ethical tension between ensuring the safety of patients and respecting their right to autonomy~\cite{burr2020ethics}. In addition, data privacy, like autonomy, is another central theme in personalised technologies—especially for digital services that rely on behavioural signals and sensitive mental health data to personalise interventions. There are a myriad of associated challenges including unintended data leakages, lack of users' technical literacy, the need of finding an appropriate balance between using less privacy-invasive monitoring and providing more tailored interventions to improve health outcomes, and so on.

We empirically investigate the multifaceted challenges of autonomy and data sharing in mental health applications from the point-of-view of users. The importance of understanding the user’s perspective stems from the fact that user disengagement represents one of the key challenges towards an improved effectiveness of digital mental health interventions ~\cite{makin2017neurocognitive, chikersal2020understanding, karapanos2015sustaining, eysenbach2005law}. Similar to pharmacological therapies, no matter how personalised and efficient a digital intervention is, a user’s adherence is a pre-requisite to receive the desired benefits. As the content in mental health applications is growing, we are likely at the dawn of expansion of personalised recommender systems~\cite{khwaja2019aligning}. Therefore, the question on how to design the user experience of delivering personalised recommendations deserves an important place in Human Computer Interaction (HCI) research. Our objective is to inform digital user experience designers on how to best promote users' engagement when providing diverse digital mental health content. To this end, we explore users’ declared preference as well as their actual app usage with respect to: 1) a primarily autonomous versus a primarily guided user experience, 2) data to be shared in order to receive recommendations. Specifically, we address the following research questions: \begin{itemize}
\item \textit{Do users prefer an autonomous or guided experience in a mental health app?}
\item \textit{Does receiving an autonomous versus guided experience impact the actual app use?}
\item \textit{To power a recommendation system, do users prefer to share smartphone data or to self-report their personality traits?}
\item \textit{Does sharing smartphone data as opposed to self-reporting personality traits influence the actual app use?}
\end{itemize}

We used a commercially available mental health application that includes more than 100 activities (i.e. interventions) and delivered it to $N=218$ participants. In a two-factor factorial design experiment, we randomly assigned half of the participants to a guided user experience and the other half to an autonomous selection of mental well-being app content. Independently, we assigned half of the participants to a self-reported way of capturing a user model and half to a consent form for sharing smartphone data--that could be used to infer the same user model. We used the Big Five personality traits~\cite{donnellan2006mini, goldberg2006international} as a user model, as personality has been widely used to personalise digital health solutions~\cite{halko2010personality} and they can be inferred passively with smartphone sensing data~\cite{chittaranjan2011s, wang2018sensing, khwaja2019modeling}. The participants were primed that they will receive personalised recommendations that are based on the data that they agreed to share. However, in reality, the recommendations were random. We opted for a random placebo experimental design, based on priming, to reduce the dependency on the recommendation system accuracy that may not always be uniform for all users, thus representing a confounding factor. Having four randomised groups allowed us to delve into the relative differences in the actual app usage and users' declared preferences as a function of the two factors—autonomy and data sharing. 

The choices put forth to mental health intervention designers are not trivial, especially in light of ethical tensions related to paternalistic design choices~\cite{floridi2016tolerant} or the possible risks arising from increasingly sensitive data streams. Yet, both design choices are important to tackle in order to unlock the value of personalised technology~\cite{floridi2018ai4people}. This paper deepens understanding of users’ preference and their actual app usage as a consequence of the app design choices, and contributes to the related debates in the HCI community and beyond.

\section{Background and Related Work}

Blom and Monk defined personalisation as "a process that increases personal relevance"~\cite{blom2000personalization}. Personalisation has gained significant attention in digital services, since providing targeted user experience has been shown to increase acceptance~\cite{jorstad2005personalization}. Particularly in health applications, personalisation was shown to increase not only engagement but also effectiveness and ultimately well-being. Noar et al~\cite{noar2007does} conducted a meta-analysis of 57 studies that used tailored messages to deliver health behaviour change interventions (for smoking cessation, diet, cancer screening, etc.), and concluded that personalised health interventions are more effective than generic ones. Zanker et al~\cite{zanker2019measuring} argued that personalisation can impact a range of outcomes including user engagement, app behaviours, and adoption rates. Recent studies have also found that personalisation of digital health apps can significantly improve health outcomes~\cite{madeira2018personalising, chawla2013bringing}, however, the manner in which personalisation is delivered to the users and how they perceived it can be even more important than the extent to which a service is really personalised~\cite{li2016does}. Our work builds on the previous literature by further exploring the topic of delivering personalised recommendations in digital mental health from the users' perspective. We explored both users' preference as well as how their engagement with the app are impacted by a) different ways of providing personalised recommendation—by giving users more or less autonomy in choosing the app content, and b) different ways of sharing the data required for delivering personalisation. Our study highlights the importance of autonomy and data privacy in the design of digital mental health services and provides key takeaways for user experience design.

\subsection{Autonomy}

Autonomy has been an important focus in HCI, and specifically in persuasive technologies. Rughinis et al.~\cite{rughinics2015touching} decoupled five dimensions of autonomy in the context of health and well-being apps including: (1) degree of control that the user has; (2) degree of functional personalisation; (3) degree of truthfulness and reliability of the information in the app; (4) users' understanding of the goal-pursuit and (5) promotion of moral values by what the app recommends. Embedding autonomy in the design of digital services impacts not only motivation and user experience but also psychological well-being. For this reason, Peters et al~\cite{peters2018designing} included autonomy as one of the three key principles in “designing for well-being” (in addition to competence and relatedness), using Self Determination Theory ~\cite{ryan2000self} as the basis for their approach. For instance, game designers have long explored the concept of autonomy and showed that the perceived autonomy in video games contributes to game enjoyment and also short-term well-being~\cite{ryan2006motivational}. While autonomy leads to improved well-being and engagement (in addition to being ethically recommended~\cite{pullman1999ethics}), providing a range of choices may act as a demotivating factor~\cite{schwartz2004paradox}. Besides, providing more guidance with tailored interventions can lead to improved effectiveness of the intervention. Hence, designers of personalised applications face conflicting requirements. 

In this study, we set to explore how the degree of autonomy impacts the users' subjective preference, as well as their engagement with a mental health application.

\subsection{Data Privacy}

Data privacy and related topics—including but not limited to transparency, control and data governance—have been extensively discussed over the past decade due to rapid technological expansion. These topics gained even more prominence after the introduction of the EU's General Data Protection Regulation (GDPR)~\cite{voigt2017eu}. The HCI community has promptly focused their efforts on understanding how these topics may impact interaction with digital services. Providing personalised recommendations typically relies on using sensitive information streams and past studies indicate that users’ attitude towards sharing potentially sensitive data was shown to be very conservative~\cite{jamal2013mining}. For mobile health apps specifically, Peng et al~\cite{peng2016qualitative} conducted six focus groups and five individual interviews with 44 participants to compare what users value the most in these kinds of apps. While participants valued the benefits of personalisation, the authors found that they were strongly hesitant to share personal information for receiving these benefits. In another study, HCI researchers conducted a “Wizard of Oz” study to investigate whether the benefits of receiving highly personalised services—Ads in particular—offsets concerns related to sharing personal data~\cite{matic2017omg}. Interestingly, the study showed that participants' concerns were less pronounced when an actual benefit of sharing the data was clearly visible. However, the users' concerns on how the system inferred the user model (concretely users’ personality) remained strongly highlighted in semi-structured interviews. On a related topic, a recent study ~\cite{kim2020understanding} explored how users perceived automatic personality detection using a mixed-methods approach. They conducted a survey (with 89 participants) to understand which data streams users were willing to share, and afterwards developed a machine learning model (with the preferred data from 32 participants) to predict personality traits. Subsequently, they interviewed 9 participants to understand how users perceived the personality prediction model after seeing the prediction results. They observed that participants' opinions on data sharing were mixed and suggested that transparency can help in addressing users’ concerns such as trust and privacy. 

In our randomised placebo study, we primed participants that the selection of recommended activities in a mental health app was personalised to their personal data. The goal was to explore if the benefits of having a personalised experience will outpower their concerns about sharing the data. The success of placebo effect was evaluated and confirmed by including a control group in the experiment. We additionally contributed to the existing literature by comparing the actual app engagement and the user’s preference towards data sharing. 

Data privacy and autonomy were emphasised as key topics in the ethics of digital well-being~\cite{floridi2018ai4people}. To the best of our knowledge our work is the first that thoroughly explores how these two elements impact users’ actual app usage and self-declared preferences in a digital mental health app. 

\section{Methods}

To understand users' preferences and the usage of a mobile mental health app in the context of delivering recommendations, we used \textit{Foundations}\footnote{https://foundations.koahealth.com/}. \textit{Foundations} was a suitable platform for our study as it contains a large library with numerous intervention activities. In this section, we detail the methodology applied in this experiment.

\subsection{Mental Health App}
\label{sec:app_description}

\begin{figure}
  \includegraphics[width=\linewidth]{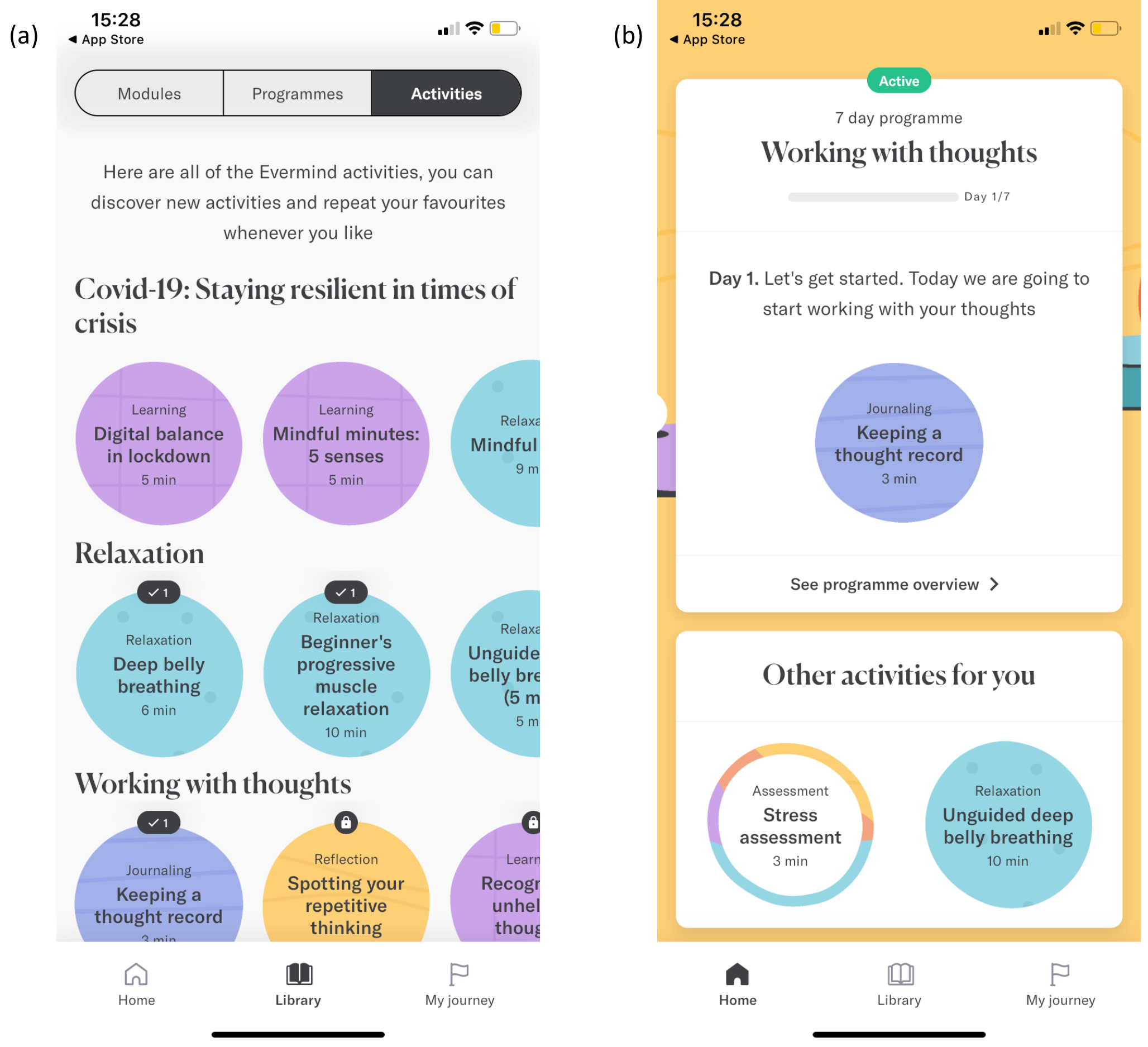}
  \caption{(a) Open library with all activities. Some activities are locked and dependent on the completion of others. (b) Recommended activities at the bottom of the home screen. All activities shown to users are randomly generated.}
  \Description[Two screenshots of the smartphone app-Foundations]{(a) Screenshot of the smartphone app-Foundations displaying colourful circles with activity titles corresponding to three different categories, namely “Covid-19: Staying resilient in times of crisis”, “Relaxation” and “Working with thoughts”. The top bar consists of free buttons (Modules, Programmes, Activities). The activities button is activated. The bottom menu shows three icons for home, library and journey. The library icon is activated.
  (b) Screenshot of the smartphone app-Foundations with a single column design and two sections. The upper section consists of the description of the “Working with thoughts” programme and a link to the programme overview. The lower section shows the heading “Other activities for you” followed by two circles with activity titles on them. The bottom menu shows three icons for home, library and journey. The home icon is activated.}
  \label{fig:Evermind}
\end{figure}

\textit{Foundations} is an evidence-based digital mental health platform designed to improve users' resilience and decrease their stress levels. At the time of this study, the version of the app incorporated 10 modules with 102 intervention activities in total. Each activity has a specific format—such as simple blog posts, relaxation audios, interactive journaling and games—to help users relax, sleep better, boost their self confidence, think positively, and similar. The app provides an open library with some activities locked in the beginning (Figure~\ref{fig:Evermind} (a)). Upon completion of each activity, users are asked to rate their experience using a thumbs up or thumbs down icon. The home screen contains a section called "Other activities for you" that shows a recommendation of two activities at a time (Figure~\ref{fig:Evermind} (b)). In our study, these recommendations were random i.e. not personalised (although presented so), which guaranteed that all the users have received the same experience. Automatic recommendations may work better for specific groups of users which would have biased the results of our study. 

\subsection{Study Design}
\label{sec:study_design}

\begin{figure*}
  \includegraphics[width=0.8\linewidth]{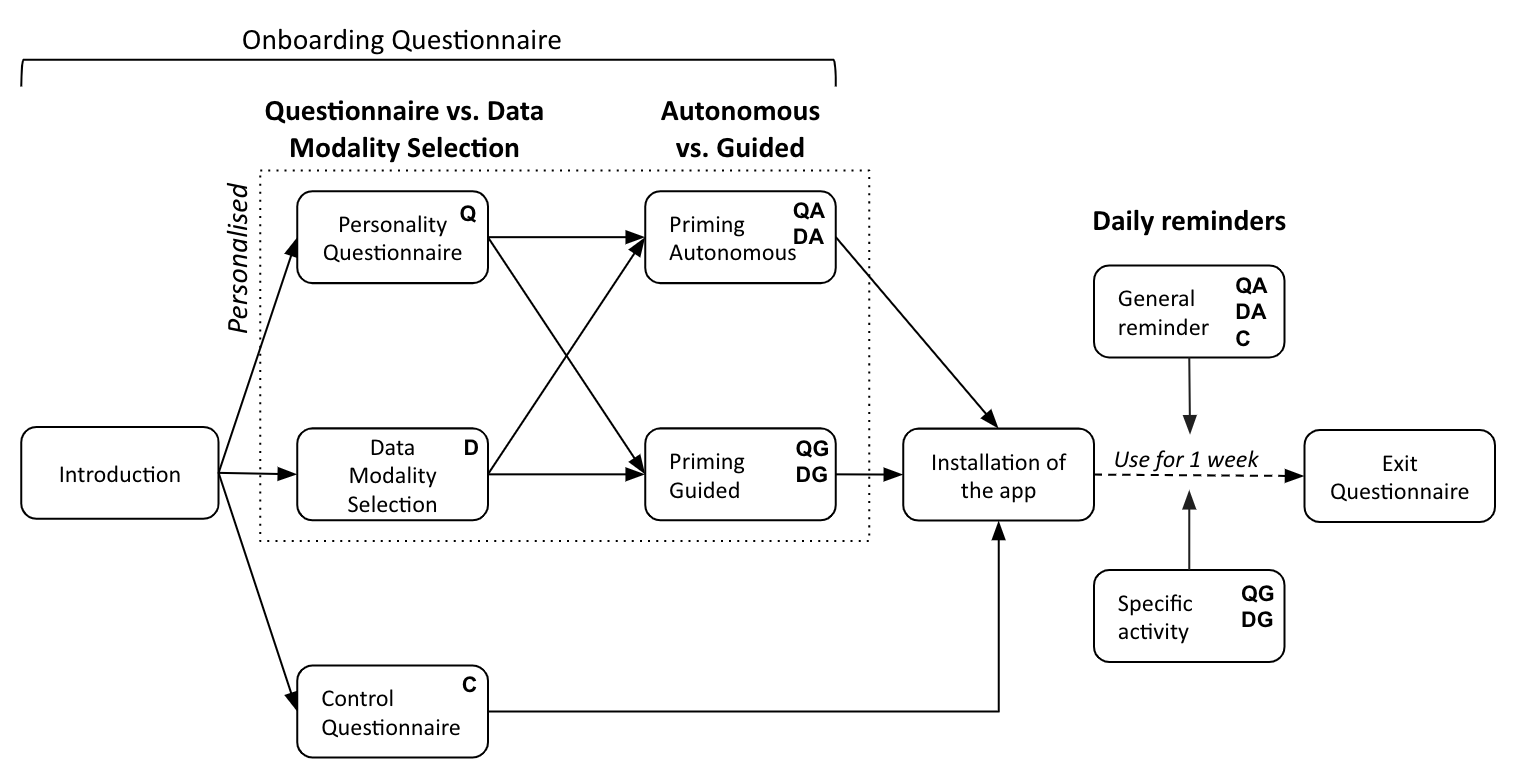}
  \caption{Experimental Design}
  \Description[Flowchart of the experimental design]{Flowchart visualising the experimental design with 10 elements and flow links. The start state is “Introduction”. The end state is “Exit questionnaire”. Details provided in section 3.2.}
  \label{fig:design}
\end{figure*}

To determine how the way of data sharing and the autonomy of the user experience impact both users' preferences and the actual usage of a mental well-being app, we designed a study consisting of three parts: (1) Onboarding questionnaire, followed by (2) the app usage for seven days with daily reminders, and finally (3) an exit questionnaire (Figure \ref{fig:design}). As the goal was to investigate the effect of the two variables, "autonomy" and "preferred way of data sharing", we designed a two-factor factorial experiment. A two-factor factorial design is an experimental design in which data is collected for all possible combinations of the levels of the two factors of interest~\cite{mukerjee2007modern}. In our case, each factor has two levels. For the preferred way of data sharing, the two levels are (1) selecting mobile sensing data and (2) completing a questionnaire, for building a personalisation model. For the former level, half of the users (randomly selected) were asked to select smartphone data streams that can be used to automatically infer their personality. The other half of the users received the 20-item personality questionnaire~\cite{donnellan2006mini} to complete. We defined two different user experiences that we refer to as "the degree of autonomy", namely (1) receiving a primarily guided user experience with the option to choose other activities out of an open library, and (2) receiving a primarily autonomous user experience with the option to use recommended activities on the home screen.

Overall, this led to 4 experimental groups that can be combined according to the variables they have in common. The combination of two groups along one identical variable is referred to as a cluster. For example, the two groups that receive an autonomous user experience—but differ in the way of data sharing—are combined and referred to as the autonomous cluster. This design allows for one group per each permutation of the two variables, which enables an analysis of all conditions separately, as well as combined. For an effect size (Cohen’s d) of 1, statistical power of 95\% and significance level of 0.05, the estimated sample size to produce a meaningful statistical significance with the Mann-Whitney test is 30. Thus, we set the criteria to have at least 30 samples in each group.

The groups differed in the onboarding questionnaire and in daily reminders during the app usage. The primary purpose of the onboarding questionnaire was to give the user the impression that the collected data will be the base for receiving personalised recommended activities in the app. However, this questionnaire was solely used for priming, and no actual personalisation was occurring in the app. All the activities participants found in the recommendation section of the app were randomly selected, as described in section \ref{sec:app_description}. The onboarding questionnaire consisted of the data sharing (smartphone modalities or questionnaire) and directions on app usage (autonomous or guided). 

Upon completion of the questionnaire, all participants received instructions on how to install \textit{Foundations} and were asked to complete at least one activity a day for one week. Daily reminders were sent according to the degree of autonomy. These reminders consisted of either a daily recommended activity for participants in the guided cluster, or a general reminder to use the app for those in the autonomous cluster. The daily recommended activities were selected from the most popular activities in the app's library. After seven days, all users completed the exit questionnaire—which was identical for all groups.

Since we did not use the users' data to personalise recommendations in the app, we included an additional control group to verify whether or not the priming was successful. 
The control group filled out a control questionnaire to match the workload to the other groups but this group did not receive any priming on personalisation.

In summary, this design resulted in having five groups:
\begin{itemize}
    \item Questionnaire-Guided \textbf{(QG)}: Personality questionnaire + daily email with activity recommendation + priming that the email recommendations are based on the reported personality
    \item Data-Guided \textbf{(DG)}: Data modality selection + daily email with activity recommendation + priming that the email recommendations are based on the automatically inferred personality
    \item Questionnaire-Autonomous \textbf{(QA)}: Personality questionnaire + daily email as a general reminder to complete one activity + priming that recommendations on the home screen are based on the reported personality
	\item Data-Autonomous \textbf{(DA)}: Data modality selection + daily email as a general reminder to complete one activity + priming that recommendations on the home screen are based on the automatically inferred personality
    \item Control \textbf{(C)}: Control questionnaire + daily email as a general reminder to complete one activity
\end{itemize}

Our study was approved by the internal ethics board. As the whole set of intervention activities in Foundations has been recently evaluated in a Randomised Control trial~\cite{catuara2021rct} and demonstrated an overall improvement in users' overall well-being, no harm was expected to be introduced by a deception study that recommends users with the most popular activities.

\subsection{Data Collection}
The onboarding and exit questionnaires were created using the Typeform~\footnote{https://www.typeform.com/} survey collection tool. We designed five variations of the onboarding questionnaire for each of the five groups defined in Section~\ref{sec:study_design}. In each of these questionnaires, participants were presented with a consent form explaining details on the data collection and purpose of the study—in compliance with the EU General Data Protection Regulation (GDPR). For users in questionnaire cluster, a 7-point Likert scale (1 strongly disagree to 7 strongly agree) was used for the personality questionnaire. Users in the data cluster were provided with 10 different smartphone sensing data categories and asked to select at least 4 that could be sampled from their smartphones. The rationale for introducing the data choice was to resemble the choice that users have in real-world applications. Android and iOS give users the possibility to opt-out from specific data streams. Moreover, in Europe—where we conducted the experiments—this is a strict regulatory requirement as per the GDPR. We selected the 10 most common sensing modalities that have been used in the previous literature to predict personality traits ~\cite{monsted2018phone, de2013predicting, chittaranjan2011s, chittaranjan2013mining, wang2018sensing, khwaja2019modeling}. The 10 options included:

\begin{itemize}
    \item Time spent with different applications (App time)
    \item Geographical location (Location)
    \item Number of steps walked (Steps)
    \item Noise in the environment (Noise)
    \item Bluetooth and WiFi data (Bluetooth/Wifi)
    \item Battery level (Battery)
    \item Ambient Light in the environment (Light)
    \item Call history (without phone numbers) (Calls)
    \item Frequency of social network usage (Social network)
    \item Phone lock/unlock data and screen usage (Un(Lock))
\end{itemize}

We asked users to select at least 4 options out of 10 and explained that selecting more options leads to a higher accuracy in inferring personality. After the onboarding, users were asked to use \textit{Foundations} for a week. App usage logs consisting of activities completed, time taken per activity etc. were recorded for each user during the study. 

Upon using the app for an entire week, the users were presented with an exit questionnaire. This questionnaire had four sections asking users \textbf{(Ex1)} about their overall experience of the mental health app and their perspectives on personalisation of the app \textbf{(Ex2)} if they prefer to have autonomy in selecting activities or have the app select the right activity for them, \textbf{(Ex3)} if they prefer to complete a personality questionnaire or provide smartphone sensing data and their privacy preferences regarding the same. Based on the Technology Acceptance Model~\cite{lee2003technology}, the first set of questions \textbf{(Ex1)} was defined to understand how users perceived the app in general. The second \textbf{(Ex2)} and third \textbf{(Ex3)} set of questions were related to the users' preference to be guided vs to have autonomy, as well as sharing the data through a questionnaire or by providing their mobile sensing data. \textbf{Ex3} also included questions related to privacy concerns (a recent study that explored personality profiling by a chatbot indicated that participants generally regarded personality as sensitive data that they would be reluctant to share~\cite{volkel2020trick}).

\textbf{Ex1} - \textbf{Ex3} were delivered as a 7-point Likert scale or a multiple choice (select 'X' or 'Y'). Additionally, we had two free text questions where the users could give suggestions on the how the app could be improved and more personalised to them (\textbf{Ex4}). Subsequently, we presented the participants with demographic questions - gender, age~\footnote{We asked range of age rather than exact number}, education level and the continent of residence. The exit questionnaire concluded with a text block that debriefed the participants.

\subsection{Participants and Inclusion Criteria}

\begin{table*}[h]
\caption{Demographics of participants}
\label{table:demographics}
\centering
\begin{tabular}{|p{1.9cm}|p{2.4cm}|p{1.3cm}|p{1.3cm}|p{1.3cm}|p{1.3cm}|p{1.3cm}|p{1.3cm}|}
\hline
Demographic     & Particular        & Complete & QG & DG & QA & DA & C\\ \hline \hline
\multicolumn{2}{|c|}{Size of Population}                   & 218      & 45                               & 52                        & 40                          & 40                    & 41          \\ \hline
Gender          & Female            & 113      & 24                               & 25                        & 21                          & 17                    & 26          \\
                & Male              & 105      & 21                               & 27                        & 19                          & 23                    & 15          \\ \hline
Age             & 15-19$^{\#}$             & 6        & -                                & 1                         & 1                           & 2                     & 2           \\
                & 20-24             & 26       & 3                                & 3                         & 8                           & 5                     & 7           \\
                & 25-29             & 23       & 6                                & 3                         & 4                           & 7                     & 3           \\
                & 30-34             & 26       & 3                                & 9                         & 4                           & 6                     & 4           \\
                & 35-39             & 25       & 7                                & 7                         & 3                           & 4                     & 4           \\
                & 40-44             & 31       & 6                                & 9                         & 5                           & 3                     & 8           \\
                & 45-49             & 17       & 4                                & 6                         & 3                           & 2                     & 2           \\
                & 50-54             & 29       & 7                                & 9                         & 5                           & 4                     & 4           \\
                & 55-59             & 11       & 4                                & 2                         & -                           & 3                     & 2           \\
                & 60+               & 24       & 5                                & 3                         & 7                           & 4                     & 5           \\\hline
Education       & Secondary School  & 69       & 14                               & 19                        & 13                          & 14                    & 9           \\
                & Bachelor's Degree & 92       & 17                               & 22                        & 17                          & 16                    & 20          \\
                & Master's Degree   & 30       & 9                                & 5                         & 4                           & 6                     & 6           \\
                & Ph.D. or higher   & 8        & 1                                & 2                         & 2                           & -                     & 3           \\
                & Trade School      & 15       & 1                                & 4                         & 4                           & 4                     & 2           \\
                & Prefer not to say & 4        & 3                                & -                         & -                           & -                     & 1       \\\hline   
\end{tabular}
\vspace{0.5ex}
\\{\small{*QG = Questionnaire-Guided, DG = Data-Guided, G3 = Questionnaire-Autonomous, G4 = Data-Autonomous, C = Control}}
\\{\small{$^{\#}$The minimum age of participants is 18. We provided this age option to maintain uniformity with the other age ranges}}
\end{table*}

The participants in our study were recruited through an external agency that operates in Europe. The inclusion criteria included a high proficiency in English and the minimal age of 18. We also required a minimum of 30 participants in each group and gender balance. In early July 2020, the recruitment agency sent an invite for the study through their internal mailing list and all the participants completed the study by the end of July 2020. All participants were recruited from Europe. Through the recruitment agency, we provided a monetary incentive to all participants who completed the study. Users were instructed that successful completion and receiving the incentive requires completing the onboarding questionnaire, installation and use of the mental health app for 1 week, and completing the exit questionnaire. Users were reminded each day that skipping any of the steps would result in their disqualification from the study. 

700 participants were registered for the study and were randomly assigned to one of the five groups. Based on the group allocation, they were asked to complete the corresponding onboarding questionnaire. All 700 users completed the onboarding questionnaire and were then instructed to install the app on their smartphones. Out of the 700 users, 353 participants installed the mental health app. For one week after installing the app, users received daily reminders to use the app and to engage with at least one activity per day. Using the app for 4 or more days qualified the users for the last stage of the study. We chose a threshold of 4 days, as anytime less than this would be insufficient to explore the app well. 241 participants fulfilled this criteria and were directed to the exit questionnaire. Finally, 218 users completed the exit questionnaire and this population was used for our analyses. The demographics of the participants are provided in Table~\ref{table:demographics}. Having more than 40 participants in each group exceeded the minimum number of completes required in each group. The demographic distribution indicates that the sample involved a diverse population.

\subsection{Statistical Methods}

To report statistics, we use the guidelines laid out in~\cite{habibzadeh2017statistical}. For normally distributed data, we report mean ($M$) and standard deviation ($SD$) and for data that deviated from the normal distribution, we report the median value ($Mdn$) and interquartile range ($IQR$). Interquartile range is defined as difference between the upper quartile (75 percentile) and lower quartlie (25 percentile). In order to compare the differences in two distributions, we use the Mann-Whitney U test (also known as the Wilcoxon rank sum test)~\cite{mcknight2010mann}. The Mann-Whitney U test is non-parametrised and works well for comparing distributions that are non-normal, as opposed to the parametric Student's t‐test. Additionally, when comparing three or more distributions, we use the Kruskal–Wallis test (the non-parametric equivalent of the one-way ANOVA)~\cite{mckight2010kruskal}. Although the experimental design would have allowed us to conduct ANOVAs (or Kruskall-Wallis tests) to look at differences between all 5 conditions, we decided not to use this statistical method because our research questions focused on degree of autonomy and data sharing separately rather than combined. The literature provided no base to hypothesise that any of those combinations could lead to significantly different preferences or behaviours and we did not want to make many pairwise comparisons only for the sake of obtaining more comparisons. 


Data processing was performed with the Python programming language. All statistical tests (except the power analysis) were conducted using the SciPy library~\cite{jones2001scipy} while data visualisation plots were generated using the Matplotlib library~\cite{hunter2007matplotlib}. The power analysis was conducted in Microsoft Excel, using the Mann-Whitney power function $MW\_POWER$ from the Real Statistics library~\cite{zaiontz2017real}.
\section{Results}

\subsection{Experimental validity}
We first tested whether the inclusion criteria and randomisation were executed according to our design. Major demographic characteristics as well as the total number of participants, were correctly balanced across the groups (Table 1). To probe the additional motivation to use the app beyond the monetary incentive, we asked participates to rate the extent to which they wanted to reduce the amount of stress levels on a Likert scale 1 to 7. The median score of 6 (\textit{IQR} = 2) suggested a generally high interest in reducing stress levels. A Kruskal-Wallis test showed no significant difference among the five groups (\textit{H}(4) = 2.34, \textit{p} > .05), which indicates that the randomisation across the groups was correctly applied and that the stress level was not expected to act as a confounding factor when comparing results across the groups.

\begin{table*}
\caption{Summary of Results}
\label{table:results}
\begin{tabular}{p{1cm}p{3.3cm}|p{4.5cm}|p{4.5cm}|}
\cline{3-4}
                                                         &                      & Autonomous vs. Guided                                              & Questionnaire vs. Data                                       \\ \hline
\multicolumn{1}{|c|}{{In-App Behaviours}} & Completed Activities & Significantly more completed activities in the autonomy cluster    & No significant difference in number of completed activities  \\ \cline{2-4}
\multicolumn{1}{|c|}{}                                  & Ratio of Recommended versus Chosen Activities     & Autonomous cluster: 25\%  \newline Guided cluster: 60\%& -                \\ \cline{2-4}
\multicolumn{1}{|c|}{}                                   & Session Duration     & No significant difference in session duration                      & No significant difference in session duration                \\ \cline{2-4} 
\multicolumn{1}{|c|}{}                                   & Activity Ratings     & Significantly higher ratings of activities in the autonomy cluster & No significant difference in activity ratings                \\ \hline
\multicolumn{1}{|c|}{{Declarative Data}}                      & Preference           & All users preferred to have a more guided user experience          & All users preferred to complete a personality questionnaire. \\ \cline{2-4} 
\multicolumn{1}{|c|}{}                                   & Privacy Preference     & -                      & All users agreed that providing mobile data had more privacy risks               \\ \hline
\multicolumn{1}{|c|}{{Onboarding Behaviours}}                      & Completion time           & -          & No significant difference in completion time \\ \hline
\end{tabular}
\end{table*}

Participants were informed that they were going to receive recommeneded activities personalised for them. However, in reality, the recommended selection of activities (both those sent daily and those included within the app) were random. Therefore, the success of our priming strategy was a prerequisite for exploring the perception and effects of personalised recommendations. Unlike other domains—such as shopping items, music, movies, etc.—where people are typically well aware of what constitutes a personalised recommendation, there is a low level of understanding of meaningful symptoms and personal characteristics when it comes to the personalisation of interventions. To this end, we compared the response to the statement “I believe that activities were personalised for me” (provided at the end of the study in the Exit questionnaire) which was rated on a scale 1-7. We compared the ratings between the personalisation cluster (QG, DG, QA and DA) and the control group; and the former rated the perceived personalisation significantly higher (\textit{U} = 2725.5, \textit{p} < .05). Despite the fact that the activity recommendations were not based on the Big Five personality traits, the participants believed so--indicating that the priming was successful.

The results from our experiment are summarised in Table~\ref{table:results} and explained in detail in the following sections. 

\subsection{Guided vs autonomous user experience}

We compare the app usage behaviours and self-reported preferences between the guided (QG+DG) and the autonomous clusters (QA+DA).

\subsubsection{App usage behaviours}
\label{app_usage_autonomy}
\begin{figure*}
  \includegraphics[width=\linewidth]{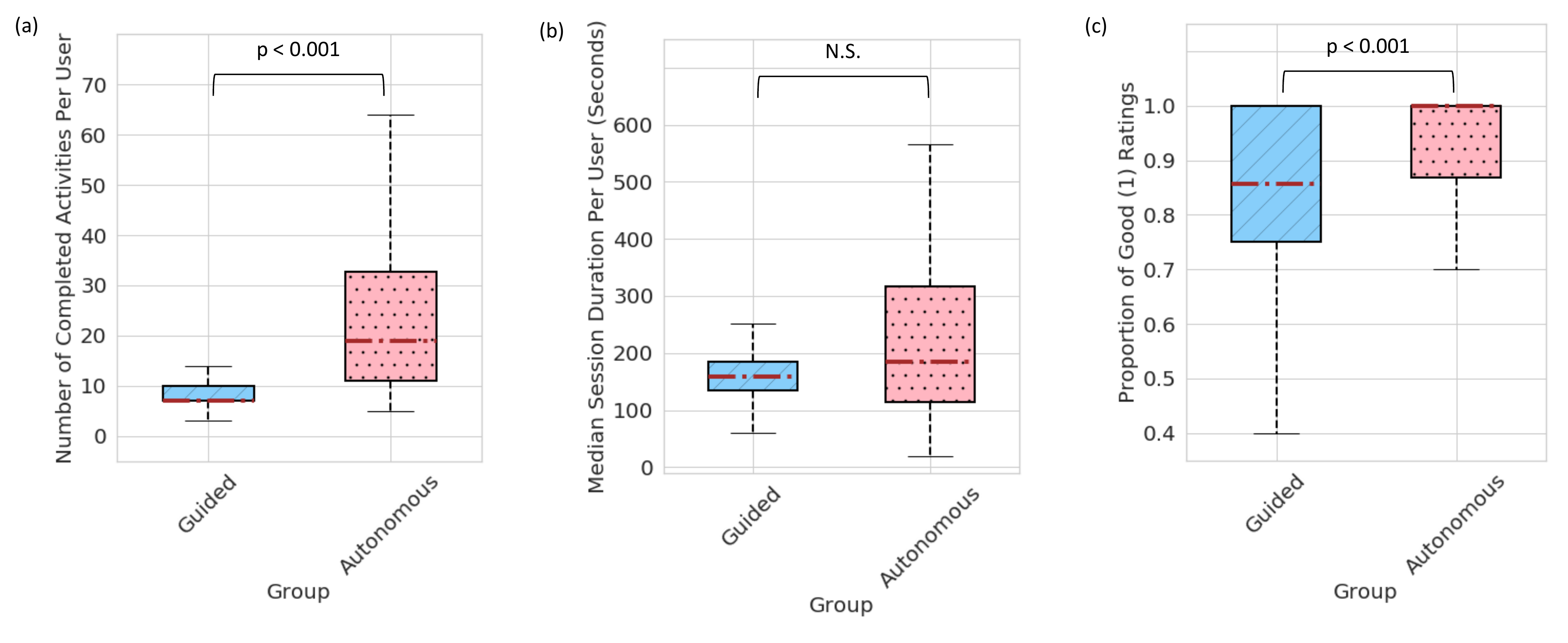}
  \caption{Differences between the guided and autonomous clusters for (a) average number of completed activities, (b) median session duration per user and (c) proportion of good (1) ratings}
  \Description[Boxplot-figures comparing app usage behaviours]{The three boxplot-figures compare the average number of completed activities, median session duration per user and the ratio of good vs bad ratings for the autonomy and guided cluster. Mean values and interquartile range are provided in section 4.2.1. The figure shows the significant difference between the two clusters for the number of completed activities and the ratio of good vs bad ratings. The median session duration per user was not significantly different between the autonomy and guided cluster. The autonomy cluster has higher median values in all three comparisons.}
  \label{fig:usage}
\end{figure*}

The number of completed activities considers only those activities that the user both started and finished. Figure~\ref{fig:usage} (a) shows that the number of activities completed by users in the autonomous cluster (\textit{Mdn} = 19, \textit{IQR} = 22.5) was significantly higher than those in the guided cluster (\textit{Mdn} = 7, \textit{IQR} = 3), \textit{U} = 1427, \textit{p} < .001. We also observed that the ratio of recommended activities from the home screen vs. voluntary chosen activities from the library amounts to 25\% for the autonomous cluster. While the ratio of recommended activities from the email reminders vs. activities from the library made up 60\% in the guided cluster. 

Subsequently, we investigated how the degree of autonomy impacted the session duration--defined as the median number of seconds for which a user was actively using the app before closing it. We observed that there was no statistical difference between autonomous (\textit{Mdn} = 184 seconds, \textit{IQR} =  363.2 seconds) and guided (\textit{Mdn} = 158 seconds, \textit{IQR} = 280.4 seconds) clusters, \textit{U} = 3346, \textit{p} > .05 (Figure~\ref{fig:usage} (b))

The design of the \textit{Foundations} provides a simple format for rating each activity, namely the users are asked to rate each activity upon its completion with either a thumbs up or thumbs down. We binary coded these ratings as 1 and 0 respectively and calculated the proportion of good (1) ratings per user--number of good ratings/(number of good+bad ratings)--which resulted in a value between 0 and 1. Figure~\ref{fig:usage} (c) shows that the proportion of good ratings of users in the autonomous cluster (\textit{Mdn} = 1, \textit{IQR} = 0.1) was significantly higher than in the guided cluster (\textit{Mdn} = 0.85, \textit{IQR} = 0.2), \textit{U} = 3047, \textit{p} < .01.

\subsubsection{Self-reported preference on autonomy}

After using the app for a week, we asked users to rate if: \textbf{A1}. They would like the mental health app to choose an activity/intervention for them (guided) and \textbf{A2}. They would like to choose an activity/intervention for themselves (autonomous). In general, users agreed more strongly that the app should provide an activity to them (\textit{Mdn} = 5, \textit{IQR} = 2), as opposed to them having autonomy to select their own activities (\textit{Mdn} = 4, \textit{IQR} = 2). The Mann-Whitney U test confirms that there is a statistical significance in their preference between the two ($U = 17051.0, p < .001$). When asked to directly compare the two options, 77.9\% of the users preferred to have an activity provided to them by the mental health app. 

Subsequently, we compared the preference for the guided and autonomous clusters separately. The percentage of users that preferred to have an activity suggested directly by the app was similar across the guided (78.4\%) and autonomous clusters (77.8\%). Next, we assessed the difference in average ratings between \textbf{A1} and \textbf{A2} within each cluster. For both the guided and autonomous clusters, users rated \textbf{A1} higher than \textbf{A2} with statistical significance ($U = 2931.5, p < .001$ and $U = 2623.5, p < .01$ respectively). This shows that, irrespective of receiving a guided or autonomous experience, all users preferred to have an app that suggests interventions for them instead of selecting activities solely on their own. 

\subsection{Questionnaire vs data selection}

We compare the app usage behaviours and self-reported preferences between the questionnaire (QG+QA) and data selection clusters (DG+DA)

\subsubsection{App usage behaviours}

Similar to the comparison described in Section~\ref{app_usage_autonomy}, we compared the number of completed activities, median session duration per user and proportion of good ratings between the questionnaire and data selection clusters. Using Mann Whitney U tests, we found no significant difference for any of these metrics (Supplementary Figure 1). 

\subsubsection{Onboarding behaviours}

We aimed to explore whether the way of data sharing (completing the personality questionnaire vs selecting the data modalities) is related to the time taken to complete the onboarding questionnaire. To do this, we compared the completion time for the questionnaire cluster against the data selection cluster. While the median time taken to complete the onboarding questionnaire was greater for the questionnaire cluster (\textit{Mdn} = 142 seconds, \textit{IQR} = 82 seconds) than the data selection cluster (\textit{Mdn} = 102 seconds, \textit{IQR} = 68 seconds), the Mann Whitney U test indicated that there was no significant difference between the two distributions ($U = 1799.5, p>.05$). The number of screens and the priming text in the onboarding questionnaires were comparable for the two clusters. The major difference in the two was the personality questionnaire versus the smartphone sensing data selections. Hence, it can be concluded that there is no significant difference between the time taken to complete the 20-item personality questionnaire and the time needed to select a subset of a list of smartphone sensing data modalities, in an onboarding process. 

\begin{figure}
  \includegraphics[width=\linewidth]{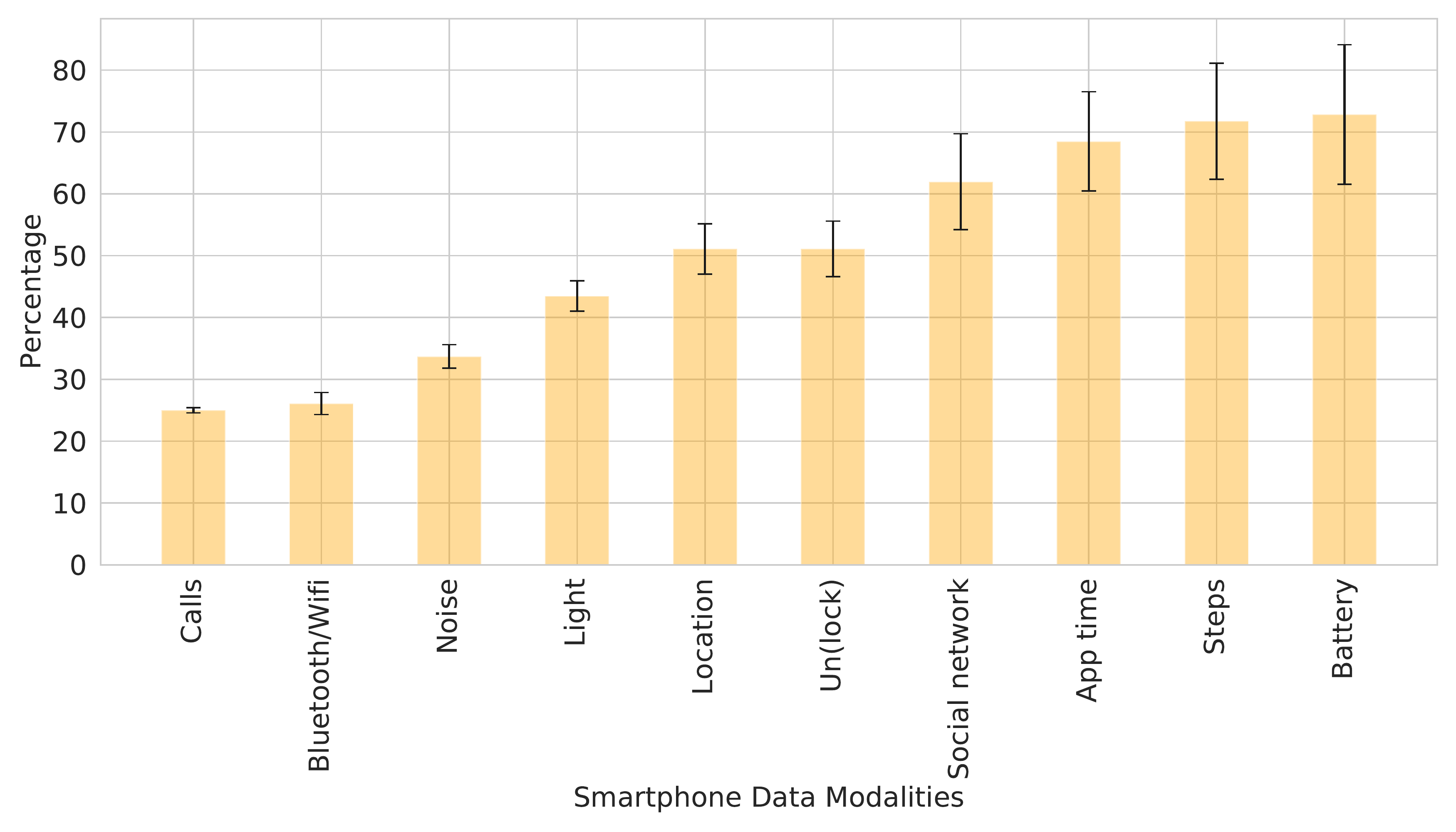}
  \caption{Proportions of users from the data sharing cluster that preferred to provide different data modalities. Column names correspond to the data modalities described in Section 3.3}
    \Description[Bar chart of the percentage of users that chose to provide each data modality]{The figure shows a bar chart visualising the percentage of users that chose to provide each data modality. The bars correspond to the data modalities described in Section 3.3. and are displayed in ascending order. The most and least selected data modalities are described in 4.3.2.}
  \label{fig:ratings}
\end{figure}

In addition, we also explored the data categories that the users in the data selection cluster were most willing to provide. Figure~\ref{fig:ratings} shows the proportion of users that provided a particular data modality. The error bars in the figure represent the standard deviation of the proportions obtained individually from DG and DA. The users were least willing to provide 1. call history (25.0\%), 2. bluetooth and wifi data (26.1\%) and 3. noise in the environment sampled from the microphone (34.0\%). As expected, these are data modalities that have the largest privacy and security concerns across both users and technologists~\cite{elkhodr2012review, mayer2016evaluating, sipior2014privacy}. Additionally, the data modalities that users are most willing to provide are 1. battery level (72.8\%), 2. number of steps walked (71.7\%) and 3. time spent on different applications (68.5\%). 


\subsubsection{Self-reported preference on data sharing}
\label{subsubsection:pref_data}

Users were asked to rate from 1 to 7: \textbf{D1}. If they were willing to complete a 5-10 min personality questionnaire (with up to 50 questions) to receive personalised recommendations for activities and \textbf{D2}. If they were willing to provide personal sensing data (e.g., GPS location) from their smartphone to receive personalised recommendations for activities. A Mann-Whitney U test confirmed with statistical significance ($U = 11568, p < .001$) that users were more willing to complete a personality questionnaire (\textit{Mdn} = 6, \textit{IQR} = 2), than provide their smartphone sensing data for personalisation (\textit{Mdn} = 4, \textit{IQR} = 3). The users were also asked to select if \textbf{D3} They would rather prefer to complete a personality questionnaire or provide their smartphone data. 90.4\% of the 218 users said they would prefer to complete a personality questionnaire to have a personalised app experience. 

Next, we compared the preferences for the questionnaire and data selection clusters. For \textbf{D3}, The percentage of users that preferred to complete the personality questionnaire instead of providing data is notably high across both the clusters (questionnaire: 92.9\% and data selection: 85.9\%). We also assessed the difference in ratings between \textbf{D1} and \textbf{D2} within each cluster. Using Mann-Whitney U tests, we observed that users in both clusters rated \textbf{D1} higher than \textbf{D2}, with statistical significance ($U = 1915, p < .001$ for the questionnaire cluster and $U = 2112.5, p < .001$ for the data selection cluster). This indicates that all users—irrespective of the way of data sharing—preferred to complete the personality questionnaire over providing their smartphone data.

\subsubsection{Self-reported preference on privacy risks}

An additional objective was to investigate if there was a difference in how users viewed privacy risks between completing a personality questionnaire and providing their smartphone data. We asked users to rate: \textbf{Pr1}. If they believed that filling out personality questionnaires for personalisation has potential privacy and data protection risks and \textbf{Pr2}. If they believed that providing a mental health app with their smartphone's sensing data for personalisation has potential privacy and data protection risks. All users believed that completing a personality questionnaire had less privacy risks (\textit{Mdn} = 4, \textit{SD} = 2) compared to providing sensing data from their smartphones (\textit{Mdn} = 5, \textit{SD} = 3). The difference between the two questions was statistically significant, $U = 21118.5, p < .05$. Within the two clusters, we also found a similar trend. Both clusters rated \textbf{Pr2} higher than \textbf{Pr1} with statistical significance ($U = 1206.5, p < .01$ for the questionnaire cluster and $U = 2106.5, p < .05$ for the data selection cluster).

\section{Discussion}

In this study, we explored how (1) the degree of autonomy in the user experience, and (2) the data to be shared impact users’ preferences and app behaviours in a mental health app. In the following, we discuss the results and highlight the main takeaways. 



\subsection{Asymmetry between in-app behaviours and preference for the degree of autonomy}

The balance between autonomy and guidance is a critical topic in personalised recommender systems, and when it comes to the area of digital mental health it has a peculiar importance. In a traditional setting, for the selection of the right intervention, autonomy is secondary to the expertise of the medical professional. However, in digital experiences, autonomy was shown to be an essential design criterion to create engagement ~\cite{peters2018designing}. Our results highlight the challenge of finding the right balance between the two and shed light on the contrast between users' preferences and their actual behaviour in the app. This together provides a set of practical takeaways for user experience designers that we discuss in the following. 

Our findings demonstrated that the difference in the degree of autonomy could influence subsequent behaviours in a mental health mobile application. We showed significant between-group differences in user behaviours, although all participants used the same application. Since there was no actual personalisation in the app, our results are independent of the accuracy of a recommendation system and solely ascribed to the perceived degree of autonomy in the user experience.

Our results challenge the popular notion that the more personalised or guided, the better an app is perceived by users. We witnessed that a primarily autonomous experience led to the greatest engagement i.e. the highest number of completed activities and best ratings. Contrary to expectations, the most guided and tailored experience appeared to discourage users' exploration and spontaneous app use. However, when asked about the subjective preference after the study had been completed, a significantly higher number of users expressed their preference for more guidance instead of autonomy. This finding shows a discrepancy between behavioural and declarative data. Our results confirm that the preferences communicated by the user do not necessarily result in quantitatively improved engagement metrics. This emphasises the importance of cautiously interpreting user research results and combining them with quantitative data, when possible, throughout the process of designing personalised user experiences.  

Interestingly, several answers to the free-text question \textit{Do you have any suggestions on how \textit{Foundations} could be more personalised for you?} referred to reminders, for instance: \textit{“Have daily reminders to help with routine”}, \textit{“Maybe a reminder to be set daily”} and \textit{“I like receiving the daily reminders. I have an 18 month old, so maybe you could set the reminder to come back on later, like a snooze button?“}. This may inspire a potential solution for an experience design that is in-between autonomy and guidance e.g. a combination of an autonomous navigation and more frequent notifications suggesting personalised content. This can result in providing more guidance without negatively impacting the users' perceived or actual agency.

In reality, none of the two clusters of users were exposed to an extreme choice between autonomy or guidance. The imposed content consumption, primarily in an autonomous versus primarily in a guided way, was clearly reflected in the actual app use—the guided cluster completed a significantly higher number of recommended activities than the autonomous cluster. However, the total number of completed activities was three times higher in the autonomous cluster. As efficacy and engagement are key pillars of digital intervention design~\cite{murray2016evaluating}, our results can be utilised by designers to optimise for these metrics. In line with our findings, the interaction in mental health apps could be designed in a similar way to popular entertainment applications such as Spotify or Netflix. Specifically, the interaction design may directly encourage autonomous navigation while providing an easy access to recommended and personalised content, thus mitigating choice overload. Moreover, different trade-offs can be made between engagement and efficacy. If the success of a specific digital therapy does not critically depend on a volume of the app use but on a targeted engagement with certain interventions, the user experience can be more guided. On the other hand, autonomous interaction designs would be more suitable to encourage a higher frequency of the app use when critical for the therapy success (e.g. meditation techniques are supposed to be practiced more regularly for optimal results).

Our results are aligned with the autonomy advocates (Ryan \& Deci \cite{deci2012self}, Peters \cite{peters2018designing}), however our findings additionally underline an important space for utilising the advantages of increasingly sophisticated recommender systems that ultimately can optimise for both efficacy and engagement.

\subsection{Users prefer questionnaires but app engagement is unaffected}

Personality traits have been used as a foundation for personalising digital health applications~\cite{halko2010personality} and for providing personalised activity recommendations that can improve mental well-being~\cite{khwaja2019aligning}. Personality traits can be obtained using questionnaires~\cite{donnellan2006mini, goldberg2006international} or inferred using machine learning models. The latter has given rise to the field of automatic personality detection. Studies in this field have shown that personality can be detected from Facebook, Twitter or Instagram usage~\cite{ferwerda2018predicting, ferwerda2018you, skowron2016fusing, hall2017say}, gaming behaviour~\cite{yee2011introverted}, music preferences~\cite{nave2018musical} and smartphone sensing data ~\cite{chittaranjan2011s, chittaranjan2013mining, de2013predicting, wang2018sensing, khwaja2019modeling}. All of these studies are based on the premise that digital behaviour data—captured passively—can be used to infer a user's personality traits automatically with machine learning, without requiring them to answer long questionnaires. However, none of these studies explored users' preferences in obtaining such data to infer a user's personality passively, especially to personalise features in a real-world application. Our work set out to answer this important question, in the context of obtaining smartphone sensing data to personalise user experience in a mental health app. 

Our results indicate that an overwhelming majority of the users prefer to complete a personality questionnaire over providing their mobile sensing data, irrespective of whether they completed the personality questionnaire before using the app or were asked to provide their smartphone data. These results are consistent with related studies showing users' improved comprehension of algorithms by using "white-box" explanations~\cite{cheng2019explaining}. Users have predominantly perceived that their smartphone sensing data entails more privacy risks than completing a personality questionnaire. This can be attributed to trust and privacy concerns with the collection of any kind of digital data~\cite{gilbert2010toward, dwyer2007trust}.

Despite the fact that smartphone sensing was perceived as obtrusive, there was no difference in app behaviour between users who completed a personality questionnaire and those who opted to provide mobile sensing data. Additionally, results from the onboarding process indicate that there is no significant difference between the time taken to complete the data consent process and the time taken to complete the 20 item personality questionnaire~\cite{donnellan2006mini}. Expectedly, users were less willing to provide more invasive data such as call history, Bluetooth data and noise from the microphone. This can have a significant impact on the accuracy of personality prediction models. Recent studies have indicated that call history data~\cite{de2013predicting}, Bluetooth data~\cite{staiano2012friends} and noise data from microphone~\cite{khwaja2019modeling, wang2018sensing} are strong predictors of personality traits. 

Should collecting mobile sensing data not be leveraged to provide other benefits to users than personality modeling for personalising the user experience, the app designers may consider avoiding the collection of smartphone data altogether. Users appear to have a strong preference towards completing a questionnaire instead and although automatic personality modelling is supposed to reduce the end user effort, it does not bring an added value in this context. This was further echoed by the users' answers to the free-text question \textit{Do you have any suggestions on how \textit{Foundations} could be more personalised for you?} including \textit{“An in depth questionnaire“}, \textit{“Maybe a regular opt-in questionnaire so you let the app know whether your conditions or state of mind is changing“} and \textit{“I think it could be more personalised by asking more about the persons life, work, family and friends.“}. This suggests that users may be willing to provide even more personal information than personality as long as they consciously and directly provide it and the app becomes more tailored to their needs as a result. As additionally suggested by the users, momentary information represents an opportunity for personalising the experience even further. In this regard, the Ecological Momentary Assessment (EMA) \cite{shiffman2008ecological} has been a widely used method that prompts users (via smartphone notifications) at different times during the day to report how they feel, what they are doing, where they are, and similar. Recent studies have shown that behaviour and mood data collected via mobile EMAs is related to mental health and health outcomes such as sleep \cite{wang2014studentlife}. Thus, data gathered from EMA surveys can point out the opportune moments to provide personalised interventions. Ultimately, the decision on gathering user models through passive sources or questionnaires requires practitioners to make a trade-off between the required amount of information, model accuracy, users' privacy concerns and a potential survey fatigue ~\cite{porter2004multiple}.

\subsection{Limitations}
Our study required us to make several trade-offs in the experimental design, which we discuss in the following.

Firstly, having identical app versions for all groups was an asset for our experimental design, although it also represented a limitation at the same time. On the one hand, it enabled us to control the perceptional aspect. On the other, having more advanced versions would have allowed us to explore the interaction between perceived accuracy and perception of personalisation, which could make the results more generalisable. 

Secondly, we did not personalise the app according to each user's actual personality which may prompt a question whether the deception of personalisation will impact the users’ trust in the app and result in a lower app usage. However, an alternative solution of providing actual personalisation would have entailed a new set of challenges. In particular, the quality of recommendations is rarely uniform and frequently biased towards specific user profiles. This issue would have been difficult or even impossible to control for. Instead, by providing random recommendations based on the most popular activities, we reduced the impact of this issue. We recognise that there is no ideal experimental design in this regard and that it entails trade-offs. However, 25\% of the completed activities in the autonomous group were recommended, which indicates that the choice of the most popular activities was appropriate. Furthermore, the recommendations were perceived as personalised, as tested between the personalisation cluster with the control group (Section 4.1).

Thirdly, we did not collect smartphone data from participants in the data group. As detailed in Section 3.3, we asked users to provide us access to their preferred data streams as a base for personalisation. However, in order not to increase the complexity of the study, we opted to use such data consent forms only as priming. Collecting smartphone sensing data would have given us an opportunity to do a more detailed behavioural analysis and further our findings. 

Lastly, all of our participants were recruited in Europe, which may have introduced a cultural bias and reduced the generalisability of our findings. 
\section{Conclusion}

In this study, we investigated how the degree of autonomy in the user experience and different ways of data sharing affect both users' preference and the actual usage of a mental well-being app. We conducted a randomised placebo study with a two-factor factorial design consisting of an onboarding questionnaire, app usage over seven days, and an exit questionnaire. 

Our results revealed an asymmetry between what users declared as their preference for autonomy (versus guidance) and how they used the app in reality. The analysis of in-app behaviours showed that a primarily autonomous design with the option to access content recommendations kept users more engaged with the app than a primarily guided experience design. However, when asked in the form of questionnaires, the majority of participants declared their preference for a more guided experience. The analysis of qualitative data suggested a potential compromise between different experience designs to satisfy both engagement metrics and subjective user preferences. 

Personalising the user experience typically requires personal data to be shared, which may impact the manner in which the app will be used. However, when analysing the actual app use, we found no impact of the data source on how users interacted with the app. Interestingly, the time taken for completing a personality questionnaire was comparable to the duration of completing a form to obtain consent for the usage of smartphone data. Yet, users indicated a strong preference for completing a personality questionnaire over providing their mobile sensing data (to infer personality). 

As mental health applications are becoming increasingly important and rich in content, our study provides key design takeaways on delivering personalised recommendations, to ultimately improve both engagement and efficacy of interventions.

\section*{Acknowledgements}
We would like to thank Emily Stott and Jordan Drewitt for their feedback and support. This work has been supported from funding awarded by the European Union's Horizon 2020 research and innovation programme, under the Marie Sklodowska-Curie grant agreement no. 722561.

\bibliographystyle{ACM-Reference-Format}
\bibliography{bibliography}


\end{document}